\documentclass[12pt]{article}
\usepackage{amsmath,amsfonts,amssymb,bbm}
\usepackage{tensor}
\usepackage[T1]{fontenc}
\usepackage[utf8]{inputenc}
\usepackage{lmodern}
\usepackage{authblk}
\usepackage[disable]{todonotes}
\usepackage{color}
\usepackage{multicol}
\usepackage{cite}
\usepackage[margin=3cm]{geometry}
\usepackage{hyperref}
\newcommand{\be}{\begin{equation}}
\newcommand{\ee}{\end{equation}}
\newcommand{\bea}{\begin{eqnarray}}
\newcommand{\eea}{\end{eqnarray}}
\newcommand{\nn}{\nonumber}

\def\crampest{\medmuskip = 1mu plus 1mu minus 1mu}

\newcommand{\n}{\nabla}

\def\cO{\mathcal{O}}
\def\cA{\mathcal{A}}

\def\cL{\mathcal{L}}

\def\be{\begin{equation}}
\def\ee{\end{equation}}
\def\benn{\begin{equation*}}
\def\eenn{\end{equation*}}
\def\beann{\begin{eqnarray*}}
\def\eeann{\end{eqnarray*}}
\def\ba{\begin{array}}
\def\ea{\end{array}}
\def\ben{\begin{enumerate}}
\def\een{\end{enumerate}}
\def\bea{\begin{eqnarray}}
\def\eea{\end{eqnarray}}
\def\p{\partial }

\advance\textheight\topskip
\DeclareFontFamily{OT1}{rsfs}{} \DeclareFontShape{OT1}{rsfs}{m}{n}{
<-7> rsfs5 <7-10> rsfs7 <10-> rsfs10}{}
\DeclareMathAlphabet{\mycal}{OT1}{rsfs}{m}{n}

\newcommand*\xbar[1]{%
  \hbox{%
    \vbox{%
      \hrule height 0.5pt % The actual bar
      \kern0.3ex%         % Distance between bar and symbol
      \hbox{%
        \kern-0.0em%      % Shortening on the left side
        \ensuremath{#1}%
        \kern-0.0em%      % Shortening on the right side
      }%
    }%
  }%
}

\newcommand\email[1]{\thanks{\href{mailto:#1}{\nolinkurl{#1}}}}

\author[a]{H. L\"u}%\email{mrhonglu@gmail.com}\,}
\author[a]{Pujian Mao}%\email{pjmao@tju.edu.cn}\,}
\author[a,b]{Jun-Bao Wu}%\email{junbao.wu@tju.edu.cn}\,}

\affil[a]{\,Center for Joint Quantum Studies and Department of Physics, School of Science, Tianjin University, 135 Yaguan Road, Tianjin 300350, China}

\affil[b]{\,Center for High Energy Physics, Peking University, 5 Yiheyuan Road, Beijing 100871, China}

\title{\bf Asymptotic Structure of Einstein-Maxwell-Dilaton Theory and Its Five Dimensional Origin\\}
\date{}

\begin{document}
\maketitle
\thispagestyle{empty}

\begin{abstract}
We consider Einstein-Maxwell-dilaton theory in four dimensions including the Kaluza-Klein theory and obtain the general asymptotic solutions in Bondi gauge. We find that there are three different types of news functions representing gravitational, electromagnetic, and scalar radiations. The mass density at any angle of the system can only decrease whenever there is any type of news function. The solution space of the Kaluza-Klein theory is also lifted to five dimensions. We also compute the asymptotic symmetries in both four dimensional Einstein-Maxwell-dilaton theory and five dimensional pure Einstein theory. We find that the symmetry algebras of the two theories are the same.
\end{abstract}

\newpage
\section{Introduction}

In 1960s, Bondi and collaborators established an elegant framework of formulating the Einstein equation as a characteristic initial value problem for axisymmetric isolated systems \cite{Bondi:1962px}. In this framework, the gravitational radiation is characterized by the news functions and the mass of the system always decreases whenever news functions exist. This demonstrates that gravitational waves exist in the full Einstein theory rather as an artifact of linearization. A surprising result of \cite{Bondi:1962px} is that they found the asymptotic symmetry group has infinite dimensions. Although all gauge choices should give the same physical result, a convenient one can make the physical properties more transparent. The manifest infinite dimensional asymptotic symmetry not only makes Bondi gauge \cite{Bondi:1962px} one of the best choices to describe physics near null infinity, but also it reveals the rich structure of spacetime in the asymptotic regions.

In recent years, physics near null infinity has obtained renewed interest from several aspects,~\textit{e.g.}~holography \cite{deBoer:2003vf,Arcioni:2003xx,Arcioni:2003td,Dappiaggi:2004kv,Dappiaggi:2005ci}, asymptotic symmetries \cite{Barnich:2009se,Barnich:2010eb,Barnich:2011ct,Barnich:2011mi,Barnich:2013axa,Donnay:2015abr}, infrared physics \cite{Strominger:2017zoo,Hawking:2016msc,Strominger:2013lka,Strominger:2013jfa,He:2014laa,
He:2014cra,He:2015zea,Conde:2016csj,Conde:2016rom,Mao:2017tey}, memory effect \cite{Strominger:2014pwa,Pasterski:2015tva,Mao:2018xcw}, and gravitational conserved quantities \cite{Godazgar:2018vmm,Godazgar:2018qpq,Godazgar:2018dvh,Kol:2019nkc,Godazgar:2019dkh,Godazgar:2019ikr}. The Bondi gauge plays a central role in all the relevant research. The asymptotic expansion of the metric functions are typically of integer powers in terms of the inverse of the radial coordinate. For extending Bondi's framework to include a matter coupled system with the same power series expansion, the matter fields are necessarily massless.
The Einstein-Maxwell theory in Bondi gauge was studied in \cite{Burg:1969,Bieri:2010tq}; however, the effect of other types of matter fields is less stressed in literatures. When spacetime dimensions are higher than four, application of the Bondi gauge is restricted. In particular, it was observed in \cite{Tanabe:2009va,Tanabe:2011es} that the news functions associated with gravitational radiation must appear in the half-integer powers of the radial expansion in five dimensions.

In the present paper, we study the asymptotic structure in a class of four dimensional Einstein-Maxwell-dilaton (EMD) theories in Bondi gauge.  The matter sector consists of Maxwell field $A$ and dilatonic scalar $\varphi$, both are massless and minimally coupled to gravity.  However, the dilaton is non-minimally coupled to the Maxwell kinetic term with an exponential function $e^{a\varphi}$ where $a$ is the dilaton coupling constant. There are many reasons to investigate EMD theory in Bondi gauge. On the one hand, this type of theories include the Kaluza-Klein theory that arises naturally from five dimensional Einstein gravity reduced on a circle.  The study of its asymptotic structure can thus provide a glimpse of that in five dimensions from the perspective of the Kaluza-Klein reduction. On the other hand, for suitable values of the constant $a$, the EMD theories can also be embedded in various supergravities that have origins in strings or M-theory. Our study of the asymptotic structure of EMD theories can thus provide a procedure to study the fundamental theories using the Bondi formalism.  Most importantly, both matter fields are massless and hence their feedback to gravity is consistent to the asymptotic integer power expansions in Bondi gauge.

The plan of the rest of this paper is quite simple. In section 2, we will study the asymptoitics of four dimensional EMD theory in detail. The solution space in Bondi gauge will be obtained where three different types of news functions are identified. The effect of the non-minimal coupling between Maxwell field and dilatonic scalar will be specified. A generalized Bondi mass-loss formula will be derived for the general EMD theories. We will work out the asymptotic symmetry group in EMD theory as well. Section 3 will turn to the study of the uplift of these solutions to solutions of five dimensional pure Einstein theory through Kaluza-Klein procedure. Asymptotic symmetry group in five dimensional Einstein theory will also be given here. We conclude this paper in the last section.

\section{Asymptotics of EMD theory in $D=4$}

\subsection{The theory}

The four-dimensional EMD theory has been extensively studied in a variety of aspects for a few decades.  The theory generalizes the Einstein-Maxwell theory to include a real dilatonic scalar. The Lagrangian is
\be\label{lagrangian}
\cL=\sqrt{-g}\left[ R - \frac{1}{4}e^{a\varphi} F^2 - \frac{1}{2} (\p \varphi)^2\right],\qquad
F=dA.
\ee
For certain specific values of the dilaton coupling constant $a$, namely $a=0,\frac1{\sqrt3},1, \sqrt3$, the EMD theory can all be embedded in the ${\cal N}=2$ STU supergravity , which is pure ${\cal N}=2$ supergravity with three vector multiplets \cite{Duff:1995sm}. The $a=0$ case can be reduced to Einstein-Maxwell theory which is the bosonic sector of ${\cal N}=2$ supergravity.  The $a=\sqrt3$ case can be Kaluza-Klein theory obtained from the circle reduction from pure gravity in five dimensions. In this section we assume that the constant $a$ is an arbitrary real constant.

The dilaton, Maxwell and Einstein equations can be derived from the Lagrangian (\ref{lagrangian}).  The covariant equations of motion are
\begin{align}
&\p_\mu (\sqrt{-g}g^{\mu\nu}\p_\nu\varphi)- \frac{a}{4} \sqrt{-g} e^{a\varphi} F^2 =0,\qquad
\p_\nu (\sqrt{-g} e^{a\varphi} F^{\mu\nu})=0,\nonumber\\
&( R_{\mu\nu} - \frac12 g_{\mu\nu} R) - \frac{1}{2} e^{a\varphi} F_{\mu\rho}{F_\nu}^\rho + \frac{1}{8} g_{\mu\nu} e^{a\varphi} F^2 - \frac{1}{2} \p_\mu \varphi \p_\nu\varphi + \frac{1}{4} g_{\mu\nu} (\p \varphi)^2 =0.
\end{align}
Contracting Eintsein equation with $g^{\mu\nu}$, one can obtain
\be\label{arrange}
R=\frac12(\p \varphi)^2.
\ee
Inserting \eqref{arrange} back, we can rearrange Einstein equation as
\be
E_{\mu\nu}\equiv R_{\mu\nu} - \frac12 e^{a\varphi} F_{\mu\rho}{F_\nu}^\rho + \frac18 g_{\mu\nu} e^{a\varphi} F^2 - \frac12 \p_\mu \varphi \p_\nu\varphi=0,\label{E}
\ee

\subsection{Bondi gauge}

We study the above EMD theory in four dimensions in Bondi gauge. The metric has the form \cite{Bondi:1962px}
\begin{multline}\label{metric}
ds^2=\left[-\frac{V(u,r,\theta)}{r}e^{2\beta(u,r,\theta)} + U(u,r,\theta)^2r^2e^{2\gamma(u,r,\theta)}\right]du^2 - 2e^{2\beta(u,r,\theta)}dudr\\
 - 2U(u,r,\theta)r^2e^{2\gamma(u,r,\theta)}dud\theta + r^2\left[e^{2\gamma(u,r,\theta)}d\theta^2 + e^{-2\gamma(u,r,\theta)}\sin^2\theta d\phi^2\right].
\end{multline}
The metric ansatz involves four functions $(V,U,\beta,\gamma)$ that are to be determined by the equations of motion.  These functions are independent of the $\phi$-coordinate and hence the metric has manifest global Killing direction $\partial_\phi$. This is the ``axisymmetric isolated system'' introduced by Bondi and collaborators \cite{Bondi:1962px}. The $g_{u\phi}$ term is noticeably absent in the metric. The inverse metric has a much simpler expression, given by
\begin{equation}
g^{\mu\nu}=\begin{pmatrix}
0 & -e^{-2\beta} & 0 & 0\\
-e^{-2\beta} & \frac{V}{r}e^{-2\beta} & -U e^{-2\beta}& 0\\
 0 & -U e^{-2\beta} & \frac{e^{-2\gamma}}{r^2} & 0\\
 0 & 0 & 0 & \frac{e^{2\gamma}}{\sin^2\theta r^2}
\end{pmatrix}.
\end{equation}
Correspondingly, we choose the following gauge fixing ansatz
\be\label{gaugefield}
A=A_u(u,r,\theta)du + A_\theta(u,r,\theta)d\theta.
\ee
{\it A priori}, we may also consider $A_\phi (u,r, \theta) d\phi$; however, we find that adding this term to the Maxwell field leads to a constraint on $A_\theta$ and $A_\phi$ from Einstein equations. The simplest solution is that $A_\phi=0$ and $A_\theta$ is an arbitrary function. We consider that this simplification is related to the metric condition $g_{u\phi}=0$ in the Bondi metric ansatz (\ref{metric}).

Following closely \cite{Bondi:1962px}, the falloff conditions for the functions $(\beta,\gamma,U,V)$ in the metric for asymptotic flatness are given by
\be\label{boundary1}
\beta=\cO(r^{-1}),\;\;\;\;\gamma=\cO(r^{-1}),\;\;\;\;U=\cO(r^{-2}),\;\;\;\;V=\cO(r).
\ee
We find that the necessary falloff conditions of the gauge and scalar fields consistent with the metric falloffs are
\be\label{boundary2}
A_u=\cO(r^{-1}),\;\;\;\;A_\theta=\cO(1),\;\;\;\;\varphi=\cO(r^{-1}).
\ee
The consistency of the Bondi gauge and the corresponding falloff conditions in the EMD theory can be verified by the equations of motion, which we carry out subsequently.

\subsection{Equations of motion in Bondi gauge}

In order to solve the equations of the EMD theory in Bondi gauge, it is useful first to rearrange the equations. Since the EMD theory \eqref{lagrangian} is a gauge theory, the equations of motion are not all independent. The constraints among them are the following identities
\be\label{identities}
\n_\mu(G^{\mu\nu}-T^{\mu\nu})=0,\;\;\;\;\;\;\p_\nu\p_\mu(\sqrt{-g} e^{a\varphi} F^{\mu\nu})=0.
\ee
Making use of these constraints, we are able to arrange the fifteen equations of motion as follows:
\begin{itemize}
\item
Five hypersurface equations:
\be\begin{split}
&\p_\nu (\sqrt{-g} e^{a\varphi} F^{u\nu})=0,\\
&E_{rr}=E_{r\theta}=E_{r\phi}=0,\\
&E_{\theta\theta}g^{\theta\theta} + E_{\phi\phi} g^{\phi\phi}=0.
\end{split}
\ee
\item
Five standard equations:
\be
\begin{split}
&\p_\nu (\sqrt{-g} e^{a\varphi} F^{\theta\nu})=\p_\nu (\sqrt{-g} e^{a\varphi} F^{\phi\nu})=0,\\
&\p_\mu (\sqrt{-g}g^{\mu\nu}\p_\nu\varphi)- \frac{a}{4} \sqrt{-g} e^{a\varphi} F^2=0,\\
&E_{\theta\theta}=E_{\theta\phi}=0.
\end{split}
\ee
\item
One trivial equation:
\be E_{ru}=0.\ee
\item
Four supplementary equations:
\be
\begin{split}
&\p_\nu (\sqrt{-g} e^{a\varphi} F^{r\nu})=0,\\
&E_{u\theta}=E_{u\phi}=E_{uu}=0.
\end{split}
\ee
\end{itemize}
As explained in the literatures \cite{Bondi:1962px,Sachs:1962wk,Barnich:2010eb,Barnich:2015jua}, once the hypersurface equations and standard equations are satisfied, the identities (\ref{identities}) yield that the trivial equation is satisfied automatically and the supplementary equations are left with only one order in the $\frac1r$ expansions\footnote{The equations that are left with only one order in the $\frac1r$ expansions due to the identity $\n_\mu(G^{\mu\nu}-T^{\mu\nu})=0$ are $G_{u\theta}-T_{u\theta}=G_{u\phi}-T_{u\phi}=G_{uu}-T_{uu}=0$. When the hypersurface equations and standard equations are satisfied, \eqref{arrange} will be guaranteed. Then $E_{u\mu}=G_{u\mu}-T_{u\mu}$.}.

\subsection{Hypersurface equations}\label{hypersurface}
Now we are ready to solve the equations of motion. Starting with $E_{rr}=0$, we obtain
\be
\p_r \beta=\frac{r}{2} (\p_r \gamma)^2 + \frac{r}{8} (\p_r \varphi)^2 + \frac{1}{8r}e^{a\varphi-2\gamma}(\p_r A_\theta)^2.
\ee
Once $\gamma$, $\varphi$ and $A_\theta$ are given, $\beta$ will be solved out.

There is only one hypersurface equation from the Maxwell's equations which is
\be\frac{1}{\sin\theta}\p_\nu (\sqrt{-g} e^{a\varphi} F^{u\nu})=0.\nn\ee
This will lead to
\be\label{maxwell}
\p_r L=\frac{1}{\sin\theta}\p_\theta\left[\sin\theta e^{a\varphi-2\gamma} \p_r A_\theta\right],
\ee
where, for later convenience, we define
\be\label{L}
L=\left(\p_r A_u + U \p_r A_\theta\right)r^2 e^{a\varphi-2\beta}.
\ee
It is completely fixed by $\gamma$, $\varphi$ and $A_\theta$.

We move on to $2 r^2 E_{r\theta}=0$, where we find
\bea
\p_r\left[r^4e^{2(\gamma-\beta)}\p_r U\right]&=&2r^2\left[\p_r\p_\theta(\beta-\gamma) + 2 \p_r\gamma\p_\theta\gamma - \frac{2\p_\theta\beta}{r} - 2\p_r\gamma \cot\theta\right]\cr
&&+ r^2 \p_r\varphi\p_\theta\varphi + L \p_r A_\theta.
\eea
To proceed, we need to implement the result of \eqref{maxwell}. Hence, $U$ will be fixed by $\beta$, $\gamma$, $\varphi$ and $A_\theta$. Then substituting $U$ back to \eqref{L}, $A_u$ can be worked out.

The next hypersurface equation is $\frac12 r^2 e^{2\beta} (E_{\theta\theta}g^{\theta\theta} + E_{\phi\phi} g^{\phi\phi})=0$, from which we have
\bea
\p_r V &=& 2r \p_\theta U+ \frac12 r^2 \p_r \p_\theta U- \frac14 r^4 e^{2(\gamma-\beta)} (\p_r U)^2+ \frac12 r^2 \p_r U \cot\theta+ 2rU\cot\theta\cr
&&+e^{2(\beta-\gamma)}\bigg[1- (\p_\theta \beta)^2- \p_\theta \beta \cot\theta + 2 \p_\theta\beta \p_\theta\gamma+ 3\p_\theta\gamma\cot\theta -2(\p_\theta\gamma)^2\cr
&& - \p^2_\theta \beta + \p^2_\theta\gamma\bigg]-\frac{1}{4r^2}L^2 e^{2\beta-a\varphi}- \frac{1}{4}  e^{2(\beta-\gamma)} (\p_\theta \varphi)^2
\eea
This will fix $V$ when $\beta$, $\gamma$, $U$, $\varphi$, $A_u$ and $A_\theta$ are known.

The last hypersurface equation $E_{r\phi}=0$ is satisfied automatically because there is no $\phi$-dependence in the un-known functions.

From hypersurface equations, we can learn that, once $\gamma$, $\varphi$ and $A_\theta$ are given as initial data, the other un-known functions $\beta$, $U$, $A_u$, and $V$ will be completely determined up to four integration constants of $r$. For the next step, we will work out the time evolutions of the initial data from the standard equations.

\subsection{Standard equations}

There are five standard equations. However two of them, namely $E_{\theta\phi}=0$ and $\p_\nu (\sqrt{-g} e^{a\varphi} F^{\phi\nu})=0$ are held automatically due to no $\phi$-dependence in our system. The rest three equations will determine the time evolution of $\gamma$, $\varphi$ and $A_\theta$, which will be calculated in this subsection.

From $\frac12 r e^{2\beta}E_{\phi\phi}g^{\phi\phi}=0$, we have
\bea\label{standardgamma}
\p_u \p_r (r\gamma)&=&\frac{1}{2r} e^{2(\beta-\gamma)}\left[1-2\p_\theta\beta\cot\theta + 3 \p_\theta\gamma\cot\theta + 2\p_\theta\beta\p_\theta\gamma - 2 (\p_\theta\gamma)^2 + \p^2_\theta\gamma\right]\cr
&&-\frac12 U \left(2\p_\theta\gamma - 3 \cot\theta + r \p_r\gamma\cot\theta + 2r \p_r\p_\theta\gamma \right)\cr
&&+\frac12 r \p_r U \cot\theta - \frac12 r \p_\theta \gamma\p_r U
- \frac{1}{2r} \p_r V + \frac{1}{2r} V \p_r \gamma\cr
 &&+ \frac12 \p_r V \p_r \gamma + \frac12 \p_\theta U - \frac12 r \p_\theta U \p_r \gamma + \frac12 V \p_r^2 \gamma\cr
&&-\frac{r}{8} e^{a\varphi-2\beta}\left[(\p_r A_u)^2 + 2 U \p_r A_u \p_r A_\theta + (U\p_r A_\theta)^2\right]\cr
 &&+ \frac{1}{8r^2} e^{a\varphi-2\gamma} \p_r A_\theta \left(2r \p_\theta A_u + V \p_r A_\theta - 2r \p_u A_\theta\right).
\eea
From $\frac{1}{2\sin\theta}e^{2\gamma-a\varphi}\p_\nu (\sqrt{-g}e^{a\varphi} F^{\theta\nu})=0$, we obtain
\bea\label{standardAtheta}
\p_u\p_r A_\theta &=& \frac12 \p_r\p_\theta A_u + \frac12 \p_r (\frac{V}{r} \p_r A_\theta)- \frac12 e^{2\gamma-a\varphi}\p_r(UL)-\frac12(a \p_u\varphi-2\p_u\gamma)\p_r A_\theta\cr
&&+ \frac12(a \p_r\varphi-2\p_r\gamma)(\frac{V}{r} \p_r A_\theta + \p_\theta A_u - \p_u A_\theta).
\eea
The last one $\frac{1}{2r\sin\theta}\left[\p_\mu (\sqrt{-g}g^{\mu\nu}\p_\nu\varphi)- \frac{a}{4} \sqrt{-g} e^{a\varphi} F^2\right]=0$ leads to
\bea\label{standardvarphi}
\p_u \p_r (r\varphi) &=& \frac{1}{2r}e^{2(\beta-\gamma)}\left[\p^2_\theta\varphi + \p_\theta\varphi\left(\cot\theta + 2 \p_\theta\beta - 2 \p_\theta\gamma\right)\right]\cr
 &&- \p_\theta\varphi U - \frac{r}{2} \p_\theta\varphi\p_r U- \frac{r}{2}\p_\theta U \p_r \varphi
-\frac{r}{2} U \left(\cot\theta\p_r\varphi + 2 \p_r \p_\theta\varphi\right)\cr
 &&+ \frac{1}{2r} V \p_r\varphi + \frac12 V \p_r^2\varphi  + \frac12 \p_r V \p_r \varphi + \frac{ar}{8}e^{a\varphi+2\beta} F^2.
\eea
Clearly, there is no constraint at the order $\cO(\frac1r)$ of $\gamma$ and $\varphi$, and at the order $\cO(1)$ of $A_\theta$ from those three equations. They are related to the news functions in the system which indicating radiations.

When the above ten equations are satisfied, $E_{ru}=0$ will be held automatically from identities (\ref{identities}).

\subsection{Solution space in series expansion}
Now, supposing that $\gamma$, $\varphi$ and $A_\theta$ are given in $\frac1r$ series expansion as initial data\footnote{The absent of the order $\cO(\frac{1}{r^2})$ in $\gamma$ is to avoid logarithm terms as explained in \cite{Bondi:1962px,Sachs:1962wk}.}
\be
\gamma=\frac{c(u,\theta)}{r} + \sum^{\infty}_{a=3}\frac{\gamma_a(u,\theta)}{r^a},
\ee
\be
\varphi=\sum^{\infty}_{a=1}\frac{\varphi_a(u,\theta)}{r^a}.
\ee
\be
A_\theta={\cA}_0(u,\theta) + \sum^{\infty}_{a=1}\frac{{\cA}_{a}(u,\theta)}{r^a},
\ee
the other functions can be worked out from the results in \ref{hypersurface}.  We find
\be
\beta=-\frac{4c^2 + \varphi_1^2 }{16r^2} - \frac{\varphi_1\varphi_2}{6 r^3}-\frac{12c\gamma_3 + 2\varphi_2^2 + 3\varphi_1\varphi_3 + \frac12 \cA_1^2}{16r^4} +\cO(r^{-5}),
\ee
\bea
U &=& -\frac{\p_\theta c + 2c \cot\theta}{r^2} + \frac{4 c \left(\p_\theta c + 2 c \cot\theta\right) - N(u,\theta )}{3 r^3}\cr
&&+\frac{1}{4r^4}\bigg[2cN - {\cA}_1q+12\cot\theta\gamma_3 -(3c^2 - \frac14 \varphi_1^2)(2c\cot\theta + \p_\theta c)\cr
&&+ 6\p_\theta\gamma_3 - \frac13\varphi_2\p_\theta \varphi_1 + \frac23\varphi_1\p_\theta \varphi_2\bigg] + \cO(r^{-5}),
\eea
\bea
A_u &=& -\frac{q(u,\theta)}{r} - \frac{{\cA}_1 \cot\theta + \p_\theta {\cA}_1-aq\varphi_1}{2r^2} + \frac{1}{24r^3} \bigg[4c^2q-8{\cA}_2 \cot\theta\cr
&& + q\varphi_1^2 - 4a^2q\varphi_1^2 + 8aq\varphi_2 + 8c\p_\theta {\cA}_1 + 4a\varphi_1\p_\theta {\cA}_1 - 8 \p_\theta {\cA}_2 \cr
&& + 4{\cA}_1(6c\cot\theta + a \cot\theta \varphi_1 + 4\p_\theta c -a \p_\theta\varphi_1)\bigg] + \cO(r^{-4}),
\eea
and
\bea
V &=& r - M(u,\theta) + \frac{1}{24r}\bigg[6q^2  + 4\p_\theta N + 4 N \cot \theta+  44 (\p_\theta c)^2\cr
&&\qquad\qquad+2c^2 \left(23+25\cos2\theta\right) \csc^2\theta + 4c \left(37\cot\theta \p_\theta c + 5 \p^2_\theta c\right)\cr
&&\qquad\qquad + 3\left( \varphi_1^2 - \cot\theta \varphi_1\p_\theta \varphi_1 + (\p_\theta\varphi_1)^2 - \varphi_1\p^2_\theta\varphi_1\right)\bigg]\cr
&&+\frac{1}{48r^2}\bigg[c^3(8-192\cot^2\theta) + 12 {\cA}_1q\cot\theta + 48 \gamma_3 - 6a q^2 \varphi_1 + 8 \varphi_1\varphi_2\cr
&& + 12 q \p_\theta {\cA}_1 + 24 N \p_\theta c + 9\cot\theta \varphi_1^2\p_\theta c - 72 \cot\theta \p_\theta\gamma_3\cr
&& - 4 \cot\theta \varphi_2\p_\theta\varphi_1 + 6\varphi_1 \p_\theta c\p_\theta\varphi_1
- 4 \cot\theta\varphi_1\p_\theta\varphi_2 + 4 \p_\theta\varphi_1 \p_\theta\varphi_2\cr
 &&+ 3\varphi_1^2\p^2_\theta c - 12 c^2(23\cot\theta \p_\theta c + \p^2_\theta c) - 24\p^2_\theta \gamma_3 \cr
&&+6c\big(8N\cot\theta - 16(\p_\theta c)^2 -(\p_\theta\varphi_1)^2 +\varphi_1(\cot\theta\p_\theta\varphi_1 + \p^2_\theta\varphi_1) - \varphi_1^2\big) \cr
&& -4\varphi_2\p^2_\theta\varphi_1 - 4\varphi_1\p^2_\theta \varphi_2\bigg]  + \cO(r^{-3}),
\eea
where $M(u,\theta)$, $N(u,\theta)$ and $q(u,\theta)$ are the integration ``constants'' from solving the partial differential equation associate with $r$.

Standard equations determine the time evolution of the whole series of $\gamma$, $\varphi$ and $A_\theta$ except for their leading order terms. In particular the first order of the standard equations are listed as follows:
\be
\p_u \varphi_2=-\frac12(\p_\theta^2\varphi_1 + \cot\theta \p_\theta\varphi_1).
\ee
\be\label{dipole}
\p_u {\cA}_1=c\p_u {\cA}_0 - \frac12\p_\theta q - \frac12 a\varphi_1 \p_u {\cA}_0.
\ee
\begin{multline}\label{grqudropole}
\p_u \gamma_3=\frac{1}{96}\bigg[c^2(16 - 32 \csc^2\theta) - 4 \cot\theta N - 3\cot\theta \p_\theta\varphi_1 \varphi_1 - 3 (\p_\theta \varphi_1)^2 + 3 \varphi_1 \p_\theta^2 \varphi_1 \\
+4c\left(6M + 3\cot\theta \p_\theta c + 5 \p_\theta^2 c \right) + 4 \left( \p_\theta N + 5 (\p_\theta c)^2 - 3 {\cA}_1 \p_u {\cA}_0 \right)\bigg].
\end{multline}
All the time evolution equations of the sub-leading terms in $\gamma$, $\varphi$ and $A_\theta$ can be derived recursively from \eqref{standardgamma}-\eqref{standardvarphi} order by order. However the time evolution of $c$, ${\cA}_0$ and $\varphi_1$ are not constrained. Hence, $\dot c$, $\dot{{\cA}}_0$ and $\dot{\varphi}_1$\footnote{An overdot denotes a time derivative $\p_u$.} are the news functions of this system that indicate gravitational, electromagnetic, and scalar radiations.

In \eqref{dipole}, the time evolution equation of $\cA_1$ involves the coupling constant $a$. Since $\cA_1$ is related to the electric dipole \cite{Janis:1965tx}, the non-minimal coupling effect can be seen from the first radiating source in the multipole expansion. On the gravitational side, the coupling constant $a$ does not show up in \eqref{grqudropole} which is related to the quadrupole \cite{Janis:1965tx}. This is a reasonable result as the scalar field is minimally coupled to gravity. Presumably, the coupling constant $a$ will show up in the time evolution equation of $\gamma_4$ which is related to the octupole.

\subsection{Conservation laws and the loss of mass}
There are four supplementary equations to be solved and we only need to solve them at one order in the $\frac1r$ expansion. Equation $E_{u\phi}=0$ holds automatically, again from the assumption that the system is $\phi$-independent. The rest three supplementary equations determine the time evolution of the integration constants $M$, $N$ and $q$. Since those integration constants are related to the conserved quantities, the supplementary equations are also called conservation equations \cite{Godazgar:2019ikr}.

From $\p_\nu (\sqrt{-g} e^{a\varphi}F^{r\nu})=0$, we obtain
\be
\p_u q =- \cot\theta \p_u  {\cA}_0 - \p_u \p_\theta {\cA}_0.
\ee
Applying the identity
\be
\oint\sin\theta (\cot\theta \p_u  {\cA}_0 + \p_u \p_\theta {\cA}_0)d\theta d\phi=2\pi \p_u  {\cA}_0 \sin\theta\mid^{\pi}_0=0,
\ee
we can conclude that the total electric charge $Q$, defined by
\be
Q=\oint q(u,\theta)\sin\theta d\theta d\phi,
\ee
is conserved. This is not surprising because the dilaton scalar field is real and it cannot carry electric charges.

The conservation law regarding the angular momentum quantity $N$ however is more subtle. The conservation equation can be obtained from $E_{u\theta}=0$. It is given by
\be
\p_u N = \p_\theta M - 3 \p_\theta c\p_u c + c \p_u\p_\theta c - \frac{3}{4} \p_u \varphi_1 \p_\theta \varphi_1 + \frac{1}{4} \varphi_1 \p_u\p_\theta \varphi_1 + q \p_u {\cA}_0.
\ee
The last supplementary equation $E_{uu}=0$ leads to
\be\label{massloss}
\p_u M=-2(\dot c)^2 - \frac{1}{2} (\dot {\cA}_0)^2 - \frac{1}{2} (\dot \varphi_1)^2 + 3 \cot\theta \p_u\p_\theta c + \p_u\p^2_\theta c - 2 \p_u c.
\ee
This is the generalized Bondi mass-loss formula in the four dimensional EMD theory. We define the mass density
\be
m=M - \frac{1}{\sin\theta}\p_\theta\left( 2\cos\theta c + \sin\theta \p_\theta c\right).
\ee
Inserting the mass density into the generalized Bondi mass-loss formula \eqref{massloss}, one obtains
\be
\p_u m=-2(\dot c)^2 - \frac{1}{2} (\dot {\cA}_0)^2 - \frac{1}{2} (\dot \varphi_1)^2.
\ee
Thus, we have the following theorem in four dimensional Einstein-Maxwell-dilaton theory:

\textit{The mass density at any angle of the system can never increase. It is a constant if and only if there is no news.}

\subsection{Asymptotic symmetries}
The complete set of local symmetry involves a pair $(\xi,\chi)$ of a vector field $\xi=\xi^\mu\p_\mu$ and an internal gauge parameter $\chi$. The generating infinitesimal transformations are given by
\be
\delta_{(\xi,\chi)}g_{\mu\nu}=\cL_{\xi}g_{\mu\nu},\;\;\;\;\delta_{(\xi,\chi)}A_\mu=\p_\mu\chi+\cL_{\xi}A_\mu,\;\;\;\;\delta_{(\xi,\chi)}\varphi=\cL_{\xi}\varphi.
\ee
The infinitesimal transformation parameters are independent of $\phi$ in order to keep the $\phi$-independence of the fields. The residual gauge transformation preserving the gauge conditions \eqref{metric} and \eqref{gaugefield} can be solved as follows:
\begin{itemize}
\item $\cL_{\xi}g_{rr}=0 \Longrightarrow \xi^u=f(u,\theta)$.
\item $\cL_{\xi}g_{r\phi}=0 \Longrightarrow \xi^\phi=\xi^\phi(u,\theta)$.
\item $\cL_{\xi}g_{r\theta}=0 \Longrightarrow g_{ru}\p_\theta f + g_{\theta\theta}\p_r\xi^\theta=0$.
\end{itemize}
The last equation can be solved as
\be
\xi^\theta = Y(u,\theta) + \int^{\infty}_r\;dr\; g_{ru}g^{\theta\theta}\p_\theta f.
\ee
\begin{itemize}
\item $\delta_{(\xi,\chi)}A_r=0 \Longrightarrow \chi=\epsilon(u,\theta) - \int^{\infty}_r\;dr\; A_\theta g^{\theta\theta}g_{ru}\p_\theta f$.
\item $\cL_{\xi}g_{u\phi}=0 \Longrightarrow \xi^\phi=\xi^\phi(\theta)$.
\item $\cL_{\xi}g_{\theta\phi}=0 \Longrightarrow \xi^\phi$ is a constant.
\end{itemize}
We have one more gauge condition from angular part of metric elements\footnote{This is equivalent to the determinant condition used in \cite{Barnich:2009se,Barnich:2010eb}. Our choice is more convenient to compare with the 5d result discussed in the next section.}
\be
\frac{ g_{\phi\phi}}{g^{\theta\theta}} =r^4\sin^2\theta.
\ee
The precise condition is $\cL_{\xi}(\frac{ g_{\phi\phi}}{g^{\theta\theta}} )=0$ which leads to
\be
\xi^r= - \frac{r}{2} \left( \p_\theta\xi^\theta + \cot\theta \xi^\theta - g^{r\theta}g_{ur} \p_\theta f \right).
\ee
Boundary conditions \eqref{boundary1} and \eqref{boundary2} will finally yield
\be
Y(u,\theta)=y \sin\theta,\;\;\;\;\epsilon(u,\theta)=\epsilon(\theta),\;\;\;\;f(u,\theta)=T(\theta) + \frac12(\p_\theta Y + \cot\theta Y)u,
\ee
where $y$ is a constant. Since there is no $\phi$ dependence in the symmetry parameters, the asymptotic symmetry group is much small than the result in \cite{Barnich:2013sxa}.

To summarize, the asymptotic symmetries of the EMD theory \eqref{lagrangian} with the gauge and boundary condition \eqref{metric}-\eqref{boundary2} are generated by
\be\begin{split}\label{4dasg1}
&\xi^u=f=T + u y \cos\theta ,\\
&\xi^r= - \frac{r}{2} \left( \p_\theta\xi^\theta + \cot\theta \xi^\theta- g^{r\theta}g_{ur} \p_\theta f \right),\\
&\xi^\theta=y \sin\theta + \p_\theta f \int^{\infty}_r \;dr\; g_{ru}g^{\theta\theta},\\
&\xi^\phi=\xi^\phi,
\end{split}
\ee
and
\be\label{4dasg2}
\chi=\epsilon -\int^{\infty}_r \;dr\; A_\theta g^{\theta\theta}g_{ru}\p_\theta f.
\ee
Notice that $\xi^r, \xi^\theta, \chi$ depend on the coupling constant $a$ through their dependence on the metric and Maxwell field.

\subsection{Asymptotic symmetry algebra}
The asymptotic symmetry transformations satisfy a modified algebra introduced in \cite{Barnich:2010eb,Barnich:2013sxa}
\be\label{4dfullalgebra}
[(\xi_1,\chi_1),(\xi_2,\chi_2)]_M=(\hat{\xi},\hat{\chi}),
\ee
where
\be
\hat{\xi}=[\xi_1,\xi_2]- \delta_{(\xi_1,\chi_1)}\xi_2 + \delta_{(\xi_2,\chi_2)}\xi_1,\;\;\;\;\;\;\hat{\chi}=\xi^\mu_1\p_\mu\chi_2-\xi^\mu_2\p_\mu\chi_1-\delta_{(\xi_1,\chi_1)}\chi_2 + \delta_{(\xi_2,\chi_2)}\chi_1
\ee
The algebra is closed which can be seen from straightforward computation
\be\begin{split}
&\hat{\xi}^u=\hat{f}= y_1 \sin\theta (\p_\theta T_2 -\cot\theta T_2)- y_2 \sin\theta (\p_\theta T_1 -\cot\theta T_1),\\
&\p_r (\hat{\xi}^\theta)=-g_{ur}g^{\theta\theta}\p_\theta \hat f,\\
&\p_r (\frac{\hat{\xi}^r}{r})=  \frac{1}{2} \left[ \p_\theta(g^{\theta\theta}g_{ur}\p_\theta \hat f) + \cot\theta (g^{\theta\theta}g_{ur}\p_\theta \hat f) + \p_r ( g^{r\theta}g_{ur} \p_\theta  \hat f)\right],\\
&\p_r (\hat{\chi})=A_\theta g^{\theta\theta}g_{ru}\p_\theta \hat f.
\end{split}\ee
When $r\rightarrow\infty$, the algebra is reduced to
\be\label{4dalgebra}
[(T_1,y_1,\epsilon_1),(T_2,y_2,\epsilon_2)]=(\hat{T},\hat{y},\hat{\epsilon}),
\ee
where
\begin{align}
&\hat{T}=  y_1 \sin\theta (\p_\theta T_2 -\cot\theta T_2)- (1\leftrightarrow2),\\
&\hat{y}=0,\\
&\hat{\epsilon}=y_1 \sin\theta \p_\theta \epsilon_2- (1\leftrightarrow2).
\end{align}
Jacobi identity for the algebra~(\ref{4dalgebra}) and the closure of the algebra~(\ref{4dfullalgebra}) guarantee that the Jacobi identity is satisfied by the algebra~(\ref{4dfullalgebra}).

To implement mode expansions, we define $t=\tan{\frac{\theta}{2}}$. In the new coordinate, we have
\begin{align}
&\hat{T}=  y_1 (t \p_t T_2 -\frac{1-t^2}{1+t^2} T_2)- (1\leftrightarrow2),\\
&\hat{y}=0,\\
&\hat{\epsilon}=y_1 t \p_t \epsilon_2- (1\leftrightarrow2).
\end{align}
The basis vectors are choosen as
\be\label{mode1}
T_m =\left(\frac{t}{1+t^2}\right)t^{m} \p_u,\;\;\;\;Y_0= t \p_t,\;\;\;\;\epsilon_m = t^{m}.
\ee
In terms of the basis vector, the asymptotic symmetry algebra is
\be\label{mode2}
[T_m,T_n]=[\epsilon_m,\epsilon_n]=[T_m, \epsilon_n]=0,
\ee
\be\label{mode3}
[Y_0,T_n]=n T_{n},\;\;\;\;[Y_0,\epsilon_n]=n \epsilon_{n}.
\ee
It is worth pointing out that the structure constants of the algebra are independent of the dilaton coupling constant $a$.  The degeneracy of the algebra is related to the restriction of the Bondi gauge where $g_{\theta\theta} \rightarrow r^2$ asymptotically, which has the effect that $Y=y\sin\theta$ rather than being a more generic function of $\theta$.

\section{Lifting to 5d}

As was discussed earlier, for specific values of the dilaton coupling constant $a$, the EMD theory can be embedded in supergravities, which implies that the order-by-order solutions we obtained in the previous section can be lifted to strings and M-theory.  It therefore provides a tool to study the more fundamental higher-dimensional theories using the Bondi formalism, via the Kaluza-Klein procedure.  In this section we shall focus on a specific example, namely $a=\sqrt3$. The theory can be obtained from $S^1$ reduction from pure Einstein gravity in five dimensions.

\subsection{Solutions}

Setting $a=\sqrt3$, the solutions of the EMD theory obtained in the previous section become those in five-dimensional Einstein gravity. The $D=5$ metric is
\be
ds^2_5=e^{-\frac{1}{\sqrt{3}}\varphi}ds^2_4 + e^{\frac{2}{\sqrt{3}}\varphi}(dz+A_\mu dx^\mu)^2.
\ee
The inverse metric is
\begin{equation}\crampest
g^{\mu\nu}=\begin{pmatrix}
0 \!&\! -e^{-2\beta+\frac{\varphi}{\sqrt3}} \!\!&\! 0 \!&\! 0 \!&\! 0 \\
-e^{-2\beta+\frac{\varphi}{\sqrt3}} \!&\! \frac{V}{r}e^{-2\beta+\frac{\varphi}{\sqrt3}} \!\!&\! -U e^{-2\beta+\frac{\varphi}{\sqrt3}} \!&\! 0 &\! e^{-2\beta+\frac{\varphi}{\sqrt3}}(A_u+A_\theta U)\\
 0 \!&\! -U e^{-2\beta+\frac{\varphi}{\sqrt3}} \!\!&\! \frac{e^{-2\gamma+\frac{\varphi}{\sqrt3}}}{r^2}
\!&\! 0 \!&\! -\frac{e^{-2\gamma+\frac{\varphi}{\sqrt3}}A_\theta}{r^2}\\
 0 \!&\! 0 \!\!&\! 0 \!&\! \frac{e^{2\gamma+\frac{\varphi}{\sqrt3}}}{r^2 \sin^2\theta } \!&\! 0\\
 0 \!&\! e^{-2\beta+\frac{\varphi}{\sqrt3}}(A_u+A_\theta U) \!&\!  -\frac{e^{-2\gamma+\frac{\varphi}{\sqrt3}}A_\theta}{r^2} \!&\! 0 \!&\! e^{-\frac{2\varphi}{\sqrt3}}+\frac{e^{-2\gamma+\frac{\varphi}{\sqrt3}}A_\theta^2}{r^2}
\end{pmatrix}
\end{equation}
Inserting the $a=\sqrt3$ solution in the previous section, we obtain $D=5$ metric as the series expansion:
\bea\label{5dmetric}
ds^2&=&-\left[1 - \frac{M+\frac{1}{\sqrt{3}}\varphi_1}{r} + \cO(r^{-2})\right]du^2\cr
&&- 2\bigg[1 - \frac{\varphi_1}{\sqrt{3}r}+ \frac{\frac12c^2 - \frac{1}{24} \varphi_1^2- \frac{\varphi_2}{\sqrt{3}}}{r^2} + \cO(r^{-3})\bigg]dudr + 2\bigg[ (2c\cot\theta + \p_\theta c)\cr
&&+ \frac{1}{3r} \left(N - 3q {\cA}_0 + 4c^2\cot\theta + c(2\p_\theta c - 2\sqrt{3} \cot\theta\varphi_1) - \sqrt{3} \varphi_1 \p_\theta c\right)\cr
&&+\cO(r^{-2}) \bigg]dud\theta -2\left[\frac{q}{r}-\frac{3{\cA}_1\cot\theta + \sqrt3 q\varphi_1 + 3 \p_\theta {\cA}_1}{6r^2} + \cO(r^{-3})\right]dudz \cr
&&
+\left[r^2 + (2c - \frac{\varphi_1}{\sqrt{3}})r + {\cA}_0^2+\frac12\Big[(\frac{\varphi_1}{\sqrt{3}}-2c)^2 - \frac{2\varphi_2}{\sqrt{3}}\Big] + \cO(r^{-1})\right]d\theta^2\cr
&& + \bigg[r^2\sin^2\theta - r\sin^2\theta(2c + \frac{\varphi_1}{\sqrt{3}})+ \frac{1}{18}\sin^2\theta\left[(6c+\sqrt{3}\varphi_1)^2 - 6\sqrt{3}\varphi_2\right]\cr
&&+ \cO(r^{-1})\bigg]d\phi^2 + 2\bigg[{\cA}_0 + \frac{{\cA}_1+\frac{2{\cA}_0\varphi_1}{\sqrt{3}}}{r}+ \frac{{\cA}_2+ \frac23[\sqrt3 {\cA}_1 \varphi_1 + {\cA}_0(\varphi_1^2 + \sqrt3\varphi_2)]}{r^2}\cr
&&+ \cO(r^{-3})\bigg]d\theta dz + \left[1 + \frac{2\varphi_1}{\sqrt{3}r} + \frac{2(\varphi_1^2 + \sqrt3\varphi_2)}{3r^2} + \cO(r^{-3})\right]dz^2.
\eea
The different types of news functions $\dot c$, $\dot{{\cA}}_0$ and $\dot{\varphi}_1$ in 4d are now purely gravitational in five dimensions. They represent gravitational radiation in five dimensions.  The extra news functions arise because the asymptotic spacetimes in five dimensions is a product of four-dimensional Minkowski spacetimes and a circle. A similar interplay happens also between 3 and 4 dimensional general relativity \cite{Ashtekar:1996cm,Ashtekar:1996cd}. The physical interpretation can be understood most clearly from a lower dimensional perspective.

\subsection{Asymptotic symmetries}
In this subsection, we derive the asymptotic symmetries in 5d pure Einstein theory. The gauge conditions which are read off from the metric \eqref{5dmetric} are
\be\label{gauge}
g_{rr}=g_{r\theta}=g_{r\phi}=g_{rz}=g_{u\phi}=g_{\theta\phi}=g_{\phi z}=0.
\ee
The infinitesimal transformation parameter $\xi^\mu$ will be independent of $\phi$ and $z$ since there is no $\phi$ nor $z$ dependence in the metric. The residual gauge transformation preserving the required gauge conditions is solved as follows:
\begin{itemize}
\item $\cL_{\xi}g_{rr}=0 \Longrightarrow \xi^u=f(u,\theta)$.
\item $\cL_{\xi}g_{r\phi}=0 \Longrightarrow \xi^\phi=\xi^\phi(u,\theta)$.
\item $\cL_{\xi}g_{r\theta}=0 \Longrightarrow g_{ru}\p_\theta f + g_{\theta\theta}\p_r\xi^\theta + g_{\theta z} \p_r \xi^z=0$.
\item $\cL_{\xi}g_{rz}=0 \Longrightarrow  g_{zz}\p_r\xi^z + g_{\theta z} \p_r \xi^\theta=0$.
\end{itemize}
The last two equations can be solved as
\begin{align}
&\xi^\theta=Y(u,\theta) + \int^\infty_r\;dr\; \frac{g_{zz}g_{ur} \p_\theta f}{g_{zz}g_{\theta\theta}-g_{\theta z}^2}=Y(u,\theta) + \int^\infty_r\;dr\; g_{ru}g^{\theta\theta}\p_\theta f,\\
&\xi^z=\epsilon(u,\theta)+ \int^\infty_r\;dr\; \frac{g_{\theta z}g_{ur} \p_\theta f}{g_{\theta z}^2 - g_{zz}g_{\theta\theta}}=\epsilon(u,\theta) +\int^\infty_r\;dr\; g_{ru}g^{z\theta}\p_\theta f.
\end{align}
\begin{itemize}
\item $\cL_{\xi}g_{u\phi}=0 \Longrightarrow \xi^\phi=\xi^\phi(\theta)$.
\item $\cL_{\xi}g_{\theta\phi}=0 \Longrightarrow \xi^\phi$ is a constant.
\item $\cL_{\xi}g_{\phi z}=0 \Longrightarrow$no more constraint as $\xi^\phi$ is independent of $z$.
\end{itemize}
We have one more gauge condition from a combination of metric elements
\be
\frac{g_{zz}g_{\phi\phi}}{g^{\theta\theta}}=r^4\sin^2\theta.
\ee
The precise condition is $\cL_{\xi}(\frac{g_{zz}g_{\phi\phi}}{g^{\theta\theta}})=0$ which leads to
\be
\xi^r= - \frac{r}{2} \left( \p_\theta\xi^\theta + \cot\theta \xi^\theta - g^{r\theta}g_{ur}\p_\theta f \right).
\ee

Now, the vector $\xi$ is fixed up to three integration constants $f(u,\theta)$, $Y(u,\theta)$ and $\epsilon(u,\theta)$. Suitable boundary conditions will further control their time evolutions. According to \eqref{5dmetric}, the boundary conditions are
\be
g_{ur}=-1+\cO(r^{-1}),\;\;\;\;g_{u\theta}=\cO(1),\;\;\;\;g_{uz}=\cO(r^{-1}),\;\;\;\;g_{\theta\theta}=r^2+\cO(r).
\ee
Those conditions will yield
\be
Y(u,\theta)=y \sin\theta,\;\;\;\;\epsilon(u,\theta)=\epsilon(\theta),\;\;\;\;f(u,\theta)=T(\theta)+\frac12(\p_\theta Y^\theta + \cot\theta Y^\theta)u.
\ee
To summarize, the asymptotic Killing vector is
\be\begin{split}
&\xi^u=f=T + \frac12(\p_\theta Y^\theta + \cot\theta Y^\theta)u,\\
&\xi^r= - \frac{r}{2} \left( \p_\theta\xi^\theta + \cot\theta \xi^\theta - g^{r\theta}g_{ur}\p_\theta f \right),\\
&\xi^\theta=Y + \p_\theta f  \int^\infty_r\;dr\; g_{ru}g^{\theta\theta},\\
&\xi^\phi=\xi^\phi,\\
&\xi^z=\epsilon + \p_\theta f \int^\infty_r\;dr\; g_{ru}g^{z\theta}.
\end{split}
\ee

\subsection{5d algebra}
The asymptotic Killing vectors will satisfy a modified algebra introduced in \cite{Barnich:2010eb}
\be
[\xi_1,\xi_2]_M=[\xi_1,\xi_2]- \delta_{\xi_1}\xi_2 + \delta_{\xi_2}\xi_1.
\ee
The algebra is closed in the sense that
\be\begin{split}
&[\xi_1,\xi_2]^u_M=\hat{f}= y_1 \sin\theta (\p_\theta T_2 -\cot\theta T_2)- y_2 \sin\theta (\p_\theta T_1 -\cot\theta T_1),\\
&\p_r ([\xi_1,\xi_2]^\theta_M)=-g_{ur}g^{\theta\theta}\p_\theta \hat f,\\
&\p_r ([\xi_1,\xi_2]^z_M)=-g_{ur}g^{\theta z}\p_\theta \hat f,\\
&\p_r ([\xi_1,\xi_2]^r_M)=  \frac{1}{2} \left[ \p_\theta(g^{\theta\theta}g_{ur}\p_\theta \hat f) + \cot\theta (g^{\theta\theta}g_{ur}\p_\theta \hat f) + \p_r (g^{r\theta}g_{ur}\p_\theta \hat f)\right].
\end{split}\ee
When $r\rightarrow\infty$, the algebra will be reduced to
\be
[(T_1,y_1,\epsilon_1),(T_2,y_2,\epsilon_2)]=(\hat{T},\hat{y},\hat{\epsilon}),
\ee
where
\begin{align}
&\hat{T}=y_1 \sin\theta (\p_\theta T_2 -\cot\theta T_2) - (1\leftrightarrow2),\\
&\hat{y}=0,\\
&\hat{\epsilon}=y_1 \sin\theta \p_\theta \epsilon_2- (1\leftrightarrow2).
\end{align}
Unsurprisingly, we recover the same algebra as \eqref{4dalgebra} in 4d EMD theory. Literally, the same mode expansion can be applied as \eqref{mode1}-\eqref{mode3}.

\section{Conclusion and discussion}
Our motivation for studying asymptotic behavior of four dimensional EMD theory is twofold: the first one is to investigate the asymptotics in cases
with coupled dynamical massless fields with various spins, the second one is to study the asymptotics of five dimensional pure gravity among a well-chosen class of solutions avoiding half-integer powers in $1/r$ expansions. The four dimensional computations were in the Bondi gauge. Three type of news functions were identified and the generalized Bondi mass-loss formula was obtained. The four dimensional solutions were uplifted to five dimensions and this gave us the guide for gauge and boundary conditions for this class of solutions to five dimensional pure Einstein theory.  Asymptotic symmetry algebras in both four and five dimensional cases were computed and they are the same. This approach of dimensional lifting also opens windows for studying more fundamental theories in even higher dimensions based on the Bondi formalism.

One of the straightforward generalizations of this work is to relax the axisymmetric condition and study the general four dimensional asymptotic flatness solutions and their uplift to five dimensions, similar to  Sachs' generalization \cite{Sachs:1962wk} of \cite{Bondi:1962px}. We do not expect any principle difficulties while the computations will be more tedious.

A more challenging point is about the asymptotic behavior of these five dimensional solutions lifted from four dimensional EMD theory when the $z$ direction is noncompact. It will be of interest to see whether the asymptotic behavior has strong dependence on the chosen null direction as what was found in \cite{Ashtekar:1996cm} in dimensional reduction from four to three dimensions. As there, we may  need to study behavior of  four dimensional  fields at timelike infinity in additional to the behavior at null infinity studied here. The studies on asymptotics of four dimensional EMD theory here also strongly motivates us to study triangular equivalent relations \cite{Strominger:2017zoo} among asymptotic symmetries, various soft theorems and memory effects in this theory. We leave these interesting questions for future studies.

\section*{Acknowledgments}
This work is supported in part by the NSFC (National Natural Science Foundation of China) Grant No.~11935009.
H.L.~is also supported in part by NSFC Grants No.~11875200 and No.~11475024. P.M.~is also supported in part by NSFC Grant No.~11905156. J.-B.W.~is also supported in part by NSFC Grants No.~11975164 and No.~11575202.

\bibliography{ref_main,ref}

\bibliographystyle{utphys}
\end{document}